\documentclass[12pt,preprint]{aastex}
\usepackage{psfig}

\begin{document}

\title{A Re-brightening of the Radio Nebula associated with the
  2004 December~27 giant flare from SGR~1806--20}

\author{J.~D.~Gelfand\altaffilmark{1},
  Y.~E.~Lyubarsky\altaffilmark{2}, D.~Eichler\altaffilmark{2},  
  B.~M.~Gaensler\altaffilmark{1}, G.~B.~Taylor\altaffilmark{3,4},
  J.~Granot\altaffilmark{3}, K.~J.~Newton-McGee\altaffilmark{5,6},  
  E.~Ramirez-Ruiz\altaffilmark{7}, C.~Kouveliotou\altaffilmark{8},
  R.~A.~M.~J.~Wijers\altaffilmark{9}}   

\altaffiltext{1}{Harvard-Smithsonian Center for Astrophysics, 60
  Garden Street, Cambridge, MA 02138}
\altaffiltext{2}{Department of Physics, Ben Gurion University, POB
  653, Be'er Sheva 84105, Israel}
\altaffiltext{3}{Kavli Institute for Particle Astrophysics and  Cosmology, Stanford University, Stanford, CA 94309}
\altaffiltext{4}{National Radio Astronomical Observatory, P.O. Box O,
  Socorro, NM 87801}
\altaffiltext{5}{School of Physics, University of Sydney, NSW 2006,
  Australia}
\altaffiltext{6}{Australia Telescope National Facility, CSIRO, P.O. Box
  76, Epping, NSW 1710, Australia}
\altaffiltext{7}{Institute for Advanced Study, Einstein Drive,
  Princeton, NJ 08540; Chandra Fellow}
\altaffiltext{8}{NASA / Marshall Space Flight Center, NSSTC, XD-12,
  320 Sparkman Drive, Huntsville, AL 35805}
\altaffiltext{9}{Astronomical Institute ``Anton Pannekoek'',
  University of Amsterdam, Amsterdam, The Netherlands}

\begin{abstract}
The 2004 Dec.~27 giant $\gamma$-ray flare detected from the magnetar
SGR~1806--20 created an expanding radio nebula which we have monitored
with the Australia Telescope Compact Array and the Very Large Array.
These data indicate that there was an increase in the observed flux
$\sim$25 days after the initial flare that lasted for $\sim$8 days,
which we believe is the result of ambient material swept-up and
shocked by this radio nebula.  For a distance to SGR~1806--20 of
15~kpc, using the properties of this rebrightening we infer that the
initial blast wave was dominated by baryonic material of mass $M \ga
10^{24.5}$~g.  For an initial expansion velocity $v\sim0.7c$ (as
derived in an accompanying paper), we infer this material had an
initial kinetic energy $E \ga 10^{44.5}$~ergs. If this material
originated from the magnetar itself, it may have emitted a burst of
ultra-high energy ($E > 1$~TeV) neutrinos far brighter than that
expected from other astrophysical sources.
\end{abstract}


\section{Introduction}
\label{intro}
The soft gamma repeater (SGR) 1806--20 is believed to be a magnetar --- a
slowly spinning isolated neutron star with an extremely high magnetic
field ($B \sim 10^{15}~{\rm gauss}$;
\citeauthor{duncan92}~\citeyear{duncan92},
\citeauthor{chryssa98}~\citeyear{chryssa98}).  On 2004~December~27, a 
giant flare of $\gamma$-rays was detected from this object
\citep{borkowski04}, only the third such event.  For a distance of 15~kpc
(\citeauthor{corbel04} \citeyear{corbel04}; \citeauthor{mcclure05}
\citeyear{mcclure05} but see \citeauthor{nature2} \citeyear{nature2}),
the Dec.~27 flare was roughly a hundred times more luminous than the
previous two such events (\citeauthor{palmer05}~\citeyear{palmer05}; 
\citeauthor{hurley05}~\citeyear{hurley05} and references
therein). Analysis of a Very Large Array (VLA) observation of 
SGR~1806--20 seven days after the flare discovered a bright, transient 
source, VLA~J180839-202439 (\citeauthor{cameron}~\citeyear{cameron}; 
\citeauthor{gaensler}~\citeyear{gaensler}), which is believed to have
been created by the magnetar during the flare.  This detection
triggered a worldwide radio monitoring effort, whose initial results 
have been presented by \citet{nature1} and by \citet{nature2}.
In particular, it has been determined that the radio source was
initially expanding with constant velocity $v\sim0.7c$ (assuming a
distance of 15~kpc and one-sided expansion) and that, after day 9, its
flux decayed as a steep power law \citep{nature1,taylor05}.

Here, we present observational evidence for a short-term rebrightening
of this radio source which we model as the result of material shocked
by ejecta from SGR~1806--20.\footnote{The dynamical properties 
  of this model are described by \citet{granot05}.}  We then fit the
observed fluxes to this model, deriving estimates for the mass and
energy of the ejecta, and discuss this model's implications for the
nature of the Dec.~27 burst.

\section{Observations and Results}
\label{obs}
As part of a long-term monitoring campaign of VLA~J180839-202439, we
have observed this source every few days with both the Australia
Telescope Compact Array (ATCA) and with the VLA.  Here we focus on
observations at 4.8~GHz from day 6 to day 63 after the outburst, as
listed in Table~\ref{fluxtab}.  The ATCA observations used a bandwidth
of 128~MHz, and SGR~1806--20 was observed for $\sim$20~minutes at this
frequency in each observation.  For each ATCA observation, we
calibrated the flux density scale using an observation of
PKS~B1934-638 at the beginning of the run, and calibrated the phase
with a short observation of PMN~J1811-2055 taken approximately every
three minutes.  To minimize background contamination, we only used
data from baselines that included the fixed antenna located
$\sim3~{\rm km}$ away from the other five antennae in the array.  The
VLA observations were reduced using the method described by
\citet{taylor05} in which final phase calibration was achieved by
self-calibrating the SGR~1806-20 data.  For both the VLA and the ATCA
observations, the radio flux density of SGR~1806--20 was measured
using by fitting the visibility data to a source whose position was a
free parameter, fitting the visibility data to a source whose position
is fixed at the location of the SGR, and measuring the peak brightness
in an image made from these visibilities.  In general, these three
methods yielded consistent results, and any differences are reflected
in the errors provided in Table~\ref{fluxtab}. 

\begin{figure}
\includegraphics[angle=90,scale=0.6]{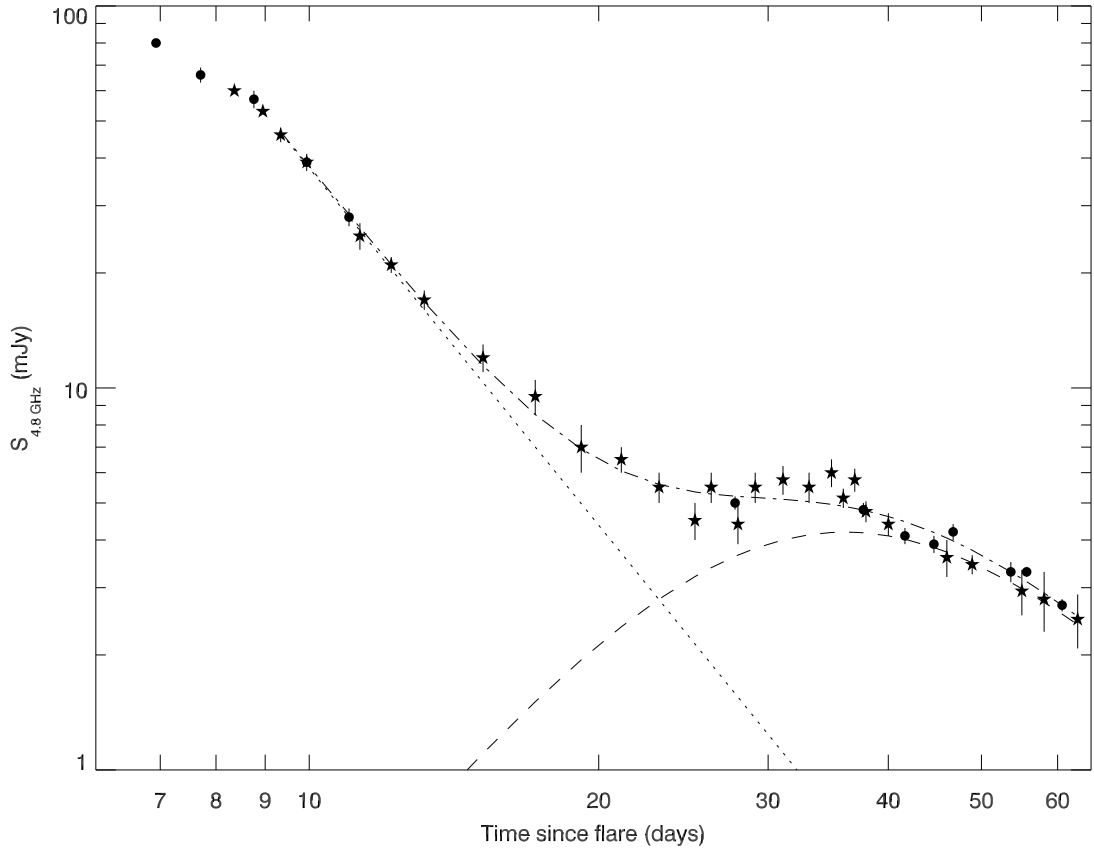}
\caption{ The 4.8~GHz light curve of the radio nebula associated with
  SGR~1806--20 up to day 62 (2005 Feb 28) after the giant flare.  The
  circles represent data taken with the VLA, and stars data taken
  with the ATCA.  The dot-dashed line in the light curve are the
  result of fitting the data to the model described in \S \ref{theory}
  and whose parameters are given in the text.  The dotted line shows
  the power-law component of the model fit while the dashed line shows
  the additional component due to the swept-up, shocked, ambient
  material.  The ends of the dot-dashed line correspond to the first
  and last data points included in the fit. \label{lcurve}} 
\end{figure}

The resultant light curve is shown in Figure \ref{lcurve}.  As reported in
\citet{nature1}, at day 9 there was a break in the light curve after
which the radio flux faded rapidly.  Starting on day 15, the
observed flux from SGR~1806--20 began to deviate significantly from a
power law decay, and on day 25 the flux began to increase for
approximately eight days.  On day 33, the observed flux began to decay
again but at a slower rate than between days 9 and 15.  In
\S\ref{theory}, we model this behavior assuming it is a result of the
source's transition from the coasting phase to the Sedov-Taylor phase
of its evolution. 

\section{A Semi-Analytic Model}
\label{theory}
In this Section, we present a semi-analytic model for the evolution of
the radio source created during the Dec.~27 giant flare.  We assume a
quasi-spherical shell of filling factor $f_b$\footnote{The results
  presented by \citet{nature1} and \citet{taylor05} suggest the radio
  source is elongated and moving along the elongation axis -- implying
  a one-sided outflow and requiring a filling factor.} and 
initial mass $M$ expanding supersonically with an initial velocity
$v_0$ into a medium of mass density $\rho$, driving a forward shock
into the ambient material. Initially, the newly swept up material is
accumulated in a thin layer between the shell and the forward shock,
and the equation of motion of this shell is: 
\begin{equation}
\frac{d}{dt} \left[ \left(M+\frac{4\pi}{3} f_b R^3\rho \right) v \right]
=4\pi f_b R^2 p,
\label{momentum}
\end{equation}
where $R=R(t)$ is the radius of the shell, $v=v(t)$ is the expansion
velocity of the shell, and $p=p(t)$, the pressure inside the shell,
found from energy conservation to be:
\begin{equation}
E\equiv \frac 12 Mv_0^2=\frac 12\left(M+\frac{4\pi}3 f_b R^3\rho\right)v^2+2\pi
f_b R^3p. \label{energy} 
\end{equation}
This approximation also works well during the Sedov-Taylor phase
(\citeauthor{zeldovich66} \citeyear{zeldovich66}), because even at
this stage most of the swept-up material is accumulated in a thin
layer just downstream of the shock whereas the rest of the volume is
filled by a rarefied, hot gas at nearly constant pressure.  By
eliminating $p$ and introducing dimensionless variables:
\begin{equation}
\label{var}
\tau \equiv \frac{t}{t_{dec}};~r \equiv \frac{R}{v_0t_{dec}};~
t_{dec} \equiv \left[ \left(\frac{4\pi f_b}{3M} \rho \right)^{1/3}v_0
  \right]^{-1}, {\rm one~finds:}
\end{equation}
\begin{equation}
 \label{difeq}
\frac{d}{d\tau}\left[\left(1+r^3\right)\frac{dr}{d\tau}\right]=\frac 1r 
\left[1-(1+r^3)\left(\frac{dr}{d\tau}\right)^2\right].
\end{equation}
At $\tau\ll 1$, the solution to Equation~(\ref{difeq}) reduces to
$v=v_0(1-0.8r^3)$.  At $\tau\gg 1$, the solution to this equation
asymptotically approaches $r=(2.5\tau^2)^{1/5}$, close to the
Sedov-Taylor solution.
 
\begin{figure}
\includegraphics[scale=0.75]{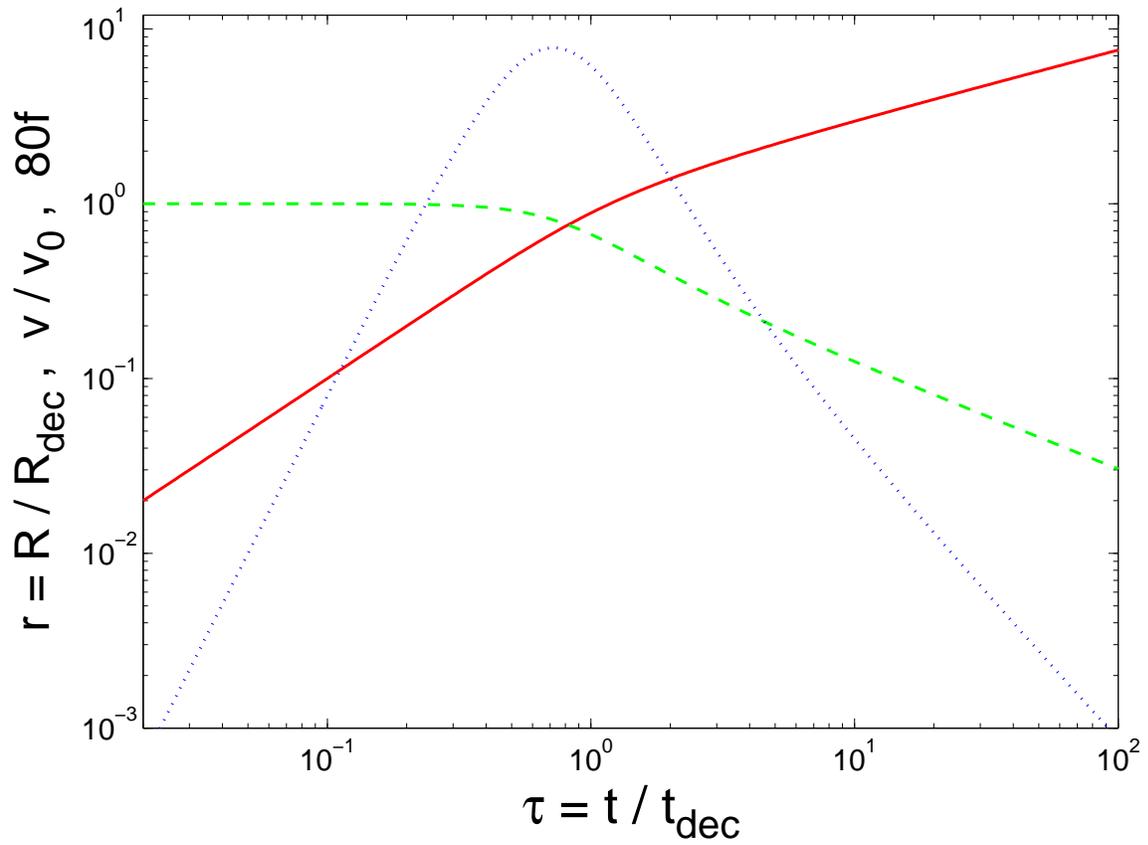}
\caption{Evolution of the expanding shell. The solid red line corresponds
  to the dimensionless radius, $r$, the dashed green line to the
  velocity $v$ in units of the initial velocity, and the dotted blue line
  corresponds to the dimensionless synchrotron flux, $f$.}\label{thepic} 
\end{figure}

We assume that, at the forward shock, electrons are heated to an
energy $\gamma_0m_ec^2=\epsilon m_pv^2$, where $\epsilon$ is proportional to
the fraction of the energy density behind the 
shock in relativistic electrons.\footnote{In the literature, the electron
  spectrum is conventionally parametrized by the fraction of the
  accelerated electrons, $\xi_e$, the fraction of the total
  energy transferred to these electrons, $\epsilon_e$, and the
  particle distribution index $p$. In order to avoid cumbersome
  expressions, we introduce $\epsilon=\epsilon_e(p-2)/2\xi_e(p-1)$.} 
Electrons with Lorentz factor $\gamma>\gamma_0$ are assumed to have a
power-law energy spectrum $N(\gamma)=K(\gamma/\gamma_0)^{-p}$ (we
assume $\gamma_0>1$, which is fulfilled for
$\epsilon>5\times10^{-4}[c/v]^2$), $N(\gamma)d\gamma$ is 
the number of electrons with energy between $\gamma m_e c^2$ and 
$(\gamma+d\gamma)m_e c^2$, $K=N(\gamma_0)$, and $p$ is the particle
distribution index -- which observationally is $p \approx 2.5$
\citep{nature1}, a typical value for shock-accelerated electrons.
Additionally, we assume that the magnetic energy density 
just downstream of the shock front is $B^2/8\pi=(9/8)\epsilon_B\rho
v^2$, where $B$ is the magnetic field strength, and $\epsilon_B$ is
the ratio of magnetic to internal energy density behind the shock.  If
the number of emitting electrons is $\sim(4\pi/3) f_b R^3\rho/m_p$,
one can estimate the emission from the swept-up material as: 
\begin{equation}
S_{\nu}=aKf_bR^3d^{-2}(\rho/m_p)\gamma_0^{p-1}B^{(1+p)/2}\nu^{(1-p)/2},
\label{flux}
\end{equation}
where $d$ is the distance to the source, $S_{\nu}$ is the flux density
at a frequency $\nu$, and $a=4.7\times 10^{-18}$ in
cgs units.  Substituting the above quantities into Equation
(\ref{flux}), one obtains:
\begin{equation}
S_{\nu}(\tau)=11 \epsilon^{1.5}_{-1} (\epsilon_{B,-1}n_{-2})^{0.87}
M_{24} v_{10}^{4.75} d_{15}^{-2} \nu_{\rm GHz}^{-0.75}
f(\tau)~{\rm mJy},\label{flux1}   
\end{equation}
where $d_{15}=d/(15~{\rm kpc})$, the ambient number density $n$ is
defined as $n\equiv\rho/m_p$ and $n_{-2}=n/(0.01~{\rm cm}^{-3})$,
$v_{10}=v/(10^{10}~{\rm cm~s}^{-1})$, $\epsilon_{-1}=\epsilon/0.1$,
$\epsilon_{B,-1}=\epsilon_B/0.1$, and the dimensionless function
$f(\tau)$ may be found from the solution $r(\tau)$ to Equation
(\ref{difeq}):
\begin{equation}
f(\tau)=r^3\left(\frac{dr}{d\tau}\right)^{(5p-3)/2}.
\end{equation}
Both $r(\tau)$ and $f(\tau)$ can be found from numerical integration
of Equation~(\ref{difeq}), and are shown in Fig. \ref{thepic}.  During
the coasting phase ($\tau \ll 1$), the luminosity grows as $t^3$ and
reaches a maximum at $\tau=0.78$, at which point the expansion velocity
has only decreased by 22\%.   At $\tau\sim$ few, the luminosity
decreases as $t^{-2}$; this is faster than the decrease during the
Sedov-Taylor phase because the pressure within the cavity remains
small for a long enough time and the expansion velocity decreases
faster than in the Sedov-Taylor solution where the expanding envelope
is filled by the hot gas.  During the Sedov phase ($\tau \ga 10$), the
luminosity decreases as $t^{-1.65}$.  However, the rate of decline
after the maximum depends strongly on the microphysics of the shock
acceleration \citep{granot05}.  We do not expect significant emission
from a reverse shock in the ejecta since it was previously shocked by
a collision with a pre-existing shell \citep{nature1,granot05}.

One can then estimate $M$ and $E$ as:
\begin{equation}
M=4.4 f_b n_{-2} t^3_{30}v_{10}^3 \times 10^{24}~{\rm g},~{\rm and}
\label{mass} 
\end{equation}
\begin{equation}
E=\frac 12 Mv_0^2=2.2 f_b n_{-2} t^3_{30}v_{10}^5 \times 10^{44}~{\rm ergs}
\label{eneb}
\end{equation}
assuming the emission peaked $30t_{30}$ days after the explosion.
This estimate for the energy is strongly dependent on $v$, whose
uncertainty is dominated by errors in the distance, not on projection
effects. If SGR~1806-20 was at a lower distance \citep{nature2}, these
estimates of $E$ and $M$ would decrease significantly, though recent
results by \citet{mcclure05} support $d\sim15$~kpc.  Additionally,
$v_{10}$ is related to $f_b$. The expansion velocity of $v\sim0.7c$~
quoted in the abstract assumes a one-sided expansion, requiring $f_b <
0.5$.  Using the elongation observed by \citet{taylor05}, we derive
$f_b \sim 0.1$.

\section{Model Fitting}
\label{analysis}
To test the model in \S\ref{theory} and to use it to independently
estimate the initial mass and energy of the source, we fit the
observed 4.8~GHz flux densities after day 8.8 to:\footnote{We only
  used data after day 8.8 in this fit because, as reported in
  \citet{nature1}, there is a break in the light curve at this epoch
  which cannot be explained by the model presented in \S\ref{theory}.} 
\begin{equation}
\label{model}
S_{\nu}(t)=S_0 \left( \frac{t}{\rm 9~days} \right)^\delta +
   11~A~\nu_{\rm GHz}^{-0.75} f(t/t_{\rm dec})~{\rm mJy}; 
\end{equation}
where $S_0$~mJy is the flux density on day 9, $\delta$ is the index of
the power-law decay, and: 
\begin{equation}
\label{aeqn}
A=\epsilon^{1.5}_{-1} (\epsilon_{B,-1}n_{-2})^{0.87} M_{24}
v_{10}^{4.75} d_{15}^{-2},
\end{equation}
as derived from Equation (\ref{flux1}).  The fit, shown in Figure
\ref{lcurve}, was performed using a minumum $\chi^2$ algorithm, and
the best-fit parameters (reduced $\chi^2=1.23$) are
$S_0=52.4\pm1.3$~mJy, $\delta=-3.12\pm0.11$, $A=11.9\pm0.2$, and
$t_{\rm dec}=46.5\pm1.7$~days.  This model predicts that at $t\approx
t_{\rm dec}$, the source's expansion velocity should decrease, as
indeed reported at this epoch by \citet{taylor05}.  The difference
between the observed and predicted shape of the rebrightening could be
due to several factors --- e.g. anisotropy in the outflow
\citep{nature1,taylor05}.  However, the fit is good enough that we
can use $A$ and $t_{\rm dec}$ to express the ejected mass in terms of
$\epsilon$, $\epsilon_{B,-1}$, $n_{-2}$, $v_{10}$ and $d_{15}$. Rather
than eliminate one of these variables, we adopt an expression for $M$
which jointly minimize the power-law dependences of all five
parameters, finding:
\begin{equation}
M = 6.6 f_b^{0.57} \epsilon^{-0.64}_{-1} \epsilon^{-0.37}_{B,-1} n^{0.20}_{-2}
v^{-0.32}_{10} d_{15}^{0.86} \times 10^{24}~{\rm g}, {\rm and}
\label{mass1}
\end{equation}
\begin{equation}
E = 3.3 f_b^{0.57} \epsilon^{-0.64}_{-1} \epsilon^{-0.37}_{B,-1} n^{0.20}_{-2}
v^{1.68}_{10} d_{15}^{0.86}~\times10^{44}~{\rm ergs}.  
\label{energy1}
\end{equation}
Here $M$ and $E$ are only weakly dependent on the ambient density, $n$
(which is difficult to constrain from observations), but are more
sensitive to the shock physics of the flow, $\epsilon$ and
$\epsilon_B$.  The total energetics of Equation (\ref{eneb}) suggest
that $n_{-2}<10^3$.  For $d_{15} \approx1$, $n_{-2} \approx 10$,
$f_b\approx0.1$ and $v_{10} \approx 2.1$ \citep{taylor05}, the
estimated initial mass is
$M=2.1\epsilon^{-0.64}_{-1}\epsilon^{-0.37}_{B,-1}\times10^{24}~{\rm
  g}$.  While $\epsilon$ and $\epsilon_B$ are unknown, we can estimate 
them from studies of gamma-ray bursts (GRBs) and supernova remnants.
If the expanding nebula behaves like the relativistic jets produced in
a GRB, then $\epsilon~\sim~10^{-2.5}-10^{-1.5}$ and
$\epsilon_B~\sim~10^{-5}-10^{-1}$ \citep{panaitescu02}, implying that
$M\sim10^{25} - 10^{27}$~g.  However, if the behavior of the expanding
nebula is closer to that of a supernova blast-wave, the magnetic field
and relativistic electrons will be in energy equipartition,
$\epsilon_B~\approx~\epsilon$ \citep{bamba03}, and
$\epsilon~\sim~10^{-2}-10^{-3}$~\citep{ellison00}, implying
$M\sim10^{26}-10^{27}$~g.  Since it is extremely unlikely that
$\epsilon$ or $\epsilon_B$ is larger than 0.1, we are rather confident
that $M \ga 2.1\times10^{24}$~g.

It is also possible that the ambient density is considerably different
from $n \approx 0.1~{\rm cm}^{-3}$.  Although the nebula initially
expanded into a cavity $\sim10^{16}$~cm in size
\citep{nature1,granot05}, by day 25 it had already expanded into the
surrounding medium.  If SGR~1806--20 is inside a stellar wind bubble
formed by its progenitor
(e.g. \citeauthor{gaensler05}~\citeyear{gaensler05}) or nearby massive
stars \citep{corbel04}, $n$ is possibly $\sim10^{-3}$~cm$^{-3}$,
implying $M\sim10^{24}$~g.  However, SGR~1806--20 is embedded in a
dust cloud, $n$ could be $\sim10~{\rm cm}^{-3}$ implying $M\sim
10^{25}$~g.  In either case, the uncertainty in $n$ does not change
the order of magnitude of the $M$ and $E$, which are similar to those
derived in Equations (\ref{mass}) and (\ref{eneb}) that depend on the
time of the peak in the light curve ($t_{30}$) but are independent of
the shock physics. As a result, we conclude that the Dec.~27 flare
created a nebula with initial mass $\ga10^{24.5}$~g and initial
kinetic energy $\ga10^{44.5}$~ergs. 

\section{Discussion}
\label{conclusion}
An inherent assumption in \S\ref{theory} is that most of the energy of
the radio source is in the form of modestly relativistic or
sub-relativistic baryons, as argued in more detail by
\citet{granot05}.  We postulate that the source of these baryons is
the neutron star itself. The giant flare is caused by, and accompanied
with, the violent restructuring of the magnetic field in which some 
magnetic field lines may, like a slingshot, throw away the matter from
the surface layers of the star. Although the canonical picture
(\citeauthor{thompson95} \citeyear{thompson95}, \citeyear{thompson01})
assumes that the magnetic stresses excite predominantly horizontal
motions of the crust, one can imagine that stretching of magnetic
field lines initially buried in the crust may break the force balance
so that some magnetic field line tubes will rise, together with the
beaded matter, into the magnetosphere. Note that the magnetic field of
$\sim 10^{15}$\ G easily overcomes the weight of the column of
$10^{14}$~g~cm$^{-2}$ so that an upper layer of width $\sim 100\;$m
may be expelled from regions with an appropriate structure of the
field. If the fraction $\zeta$ of the giant flare energy,
$E=10^{46}E_{46}$ erg, is transferred to the ejected matter, a mass of
$10^{26}\zeta E_{46}$~g may be ejected.   As discussed in
\citet{granot05}, if all of the inferred ejecta were released from the
surface of the NS during the initial ``hard spike'' ($\la 0.5\;$s) of
the giant flare, the outflow would be opaque to $\gamma$-rays and the 
Dec.~27 flare would not have been observed. This can be avoided if
there are regions on the magnetar surface from which radiation is
expelled without matter, and other points from which matter is expelled. 

One possible observational signature of this process is the detection
of ultra high-energy (UHE; $E_{\nu}>1$~TeV) neutrinos from SGR~1806--20
coincident with the Dec.~27 flare.  In this non-relativistic wind,
internal shocks produced by significant variations in the outflow velocity
within 0.5 light seconds of the star will accelerate some protons to
energies high enough that they create pions through collisions with
other protons.  When these pions decay, they can produce TeV
neutrinos.   If the total energy in neutrinos is $\epsilon_{\nu} E$,
where $E$ is the initial kinetic energy of the ejecta as estimated in
Equations (\ref{eneb}) and (\ref{energy1}), then the observed fluence
of neutrinos, ${\cal F}_{\nu}$, is: 
\begin{equation}
{\cal F}_{\nu} \approx 1.2 \epsilon_{\nu,-1} E_{44.5} d_{15}^{-2}
\times 10^{-3}~{\rm ergs~cm^{-2}} 
\end{equation}
where $E_{44.5}=E/10^{44.5}$~{ergs} and $\epsilon_{\nu,-1}=\epsilon_{\nu}/0.1$ 
(e.g. \citeauthor{eichler78} \citeyear{eichler78}).  If
$\epsilon_{\nu}\sim0.1$, this is much higher than the
$10^{-5}$~erg~cm$^{-2}$ typically expected from bright GRBs
\citep{eichler94}.  Depending on the exact values of $\epsilon_{\nu}$
and $E$, these neutrinos could possibly have been detected with 
current arrays, and the Dec.~27 event thus makes the best test case 
so far for testing the hypothesis of UHE neutrino emission from
 $\gamma$-ray outbursts.  It is not expected than any UHE neutrinos
will be produced in the forward shock generated by the outflow as it
expands into the ISM \citep{fan05}.

\acknowledgements
NRAO is a facility of the NSF operated under cooperative agreement by
AUI.  The ATCA is funded by the Commonwealth of Australia for
operation as a National Facility managed by CSIRO.  We thank  Bob
Sault, Barry Clark, and Joan Wrobel for scheduling the observations,
and Roland Crocker and John Raymond for useful conversations.  The
authors acknowledge the support of NASA, the Israel-U.S. BSF, the ISF,
the German-Israeli Foundation, and the U.S. DOE.

\bibliographystyle{apj}
\bibliography{burst}

\clearpage

\begin{deluxetable}{ccc|c}
\tabletypesize{\scriptsize}
\tablewidth{0pt}
\tablecolumns{4}
\tablecaption{Radio Observations at 4.8~GHz of the radio nebula produced by SGR
  1806--20\label{fluxtab}}
\tablehead{
\colhead{Average Epoch [UT]} & \colhead{Days after Burst} & 
\colhead{Telescope} & \colhead{$S_{4.8~{\rm GHz}}$ [mJy]}}
\startdata
2005~Jan~03.83 & 6.93 & VLA & 80$\pm$1 \\
2005~Jan~04.61 & 7.71 & VLA & 66$\pm$3 \\
2005~Jan~05.26 & 8.36 & ATCA & 60$\pm$1 \\
2005~Jan~05.66 & 8.76 & VLA & 57$\pm$3 \\
2005~Jan~05.85 & 8.95 & ATCA & 53$\pm$1 \\
2005~Jan~06.24 & 9.34 & ATCA & 46$\pm$2 \\
2005~Jan~06.84 & 9.94 & ATCA & 39$\pm$2 \\
2005~Jan~06.84 & 9.94 & VLA & 39$\pm$1 \\
2005~Jan~07.90 & 11.00 & VLA & 28$\pm$2 \\
2005~Jan~08.19 & 11.29 & ATCA & 25$\pm$2 \\
2005~Jan~09.07 & 12.17 & ATCA & 21$\pm$1 \\
2005~Jan~10.07 & 13.17 & ATCA & 17$\pm$1 \\
2005~Jan~12.06 & 15.16 & ATCA & 12$\pm$1 \\
2005~Jan~14.08 & 17.18 & ATCA & 10$\pm$1 \\
2005~Jan~16.08 & 19.18 & ATCA & 7$\pm$1 \\
2005~Jan~18.01 & 21.11 & ATCA & 6.5$\pm$0.5 \\
2005~Jan~20.01 & 23.11 & ATCA & 5.5$\pm$0.5 \\
2005~Jan~22.08 & 25.18 & ATCA & 4.5$\pm$0.5 \\
2005~Jan~23.08 & 26.18 & ATCA & 5.5$\pm$0.5 \\
2005~Jan~24.62 & 27.72 & VLA & 5.0$\pm$0.2 \\
2005~Jan~24.81 & 27.91 & ATCA & 4.4$\pm$0.5 \\
2005~Jan~25.99 & 29.09 & ATCA & 5.5$\pm$0.5 \\
2005~Jan~27.99 & 31.09 & ATCA & 5.8$\pm$0.5 \\
2005~Jan~29.99 & 33.09 & ATCA & 5.5$\pm$0.5 \\
2005~Jan~31.82 & 34.92 & ATCA & 6.0$\pm$0.5 \\
2005~Feb~01.82 & 35.92 & ATCA & 5.2$\pm$0.3 \\
2005~Feb~02.82 & 36.92 & ATCA & 5.8$\pm$0.4 \\
2005~Feb~03.59 & 37.69 & VLA & 4.8$\pm$0.2 \\
2005~Feb~03.82 & 37.92 & ATCA & 4.8$\pm$0.3 \\
2005~Feb~05.91 & 40.01 & ATCA & 4.4$\pm$0.3 \\
2005~Feb~07.53 & 41.63 & VLA & 4.1$\pm$0.2 \\
2005~Feb~10.52 & 44.62 & VLA & 3.9$\pm$0.2 \\
2005~Feb~11.92 & 46.02 & ATCA & 3.6$\pm$0.4 \\
2005~Feb~12.62 & 46.72 & VLA & 4.2$\pm$0.2 \\
2005~Feb~14.80 & 48.90 & ATCA & 3.5$\pm$0.2 \\
2005~Feb~19.54 & 53.64 & VLA & 3.3$\pm$0.2 \\
2005~Feb~20.98 & 55.08 & ATCA & 2.9$\pm$0.4 \\
2005~Feb~21.61 & 55.71 & VLA & 3.3$\pm$0.1 \\
2005~Feb~23.99 & 58.09 & ATCA & 2.8$\pm$0.5 \\
2005~Feb~26.55 & 60.65 & VLA & 2.7$\pm$0.1 \\
2005~Feb~28.85 & 62.95 & ATCA & 2.5$\pm$0.4 \\
\enddata
\tablecomments{Flux densities before 2005~Jan 18.01 are also reported in
  the Supplementary Section of \citet{nature1}.}  
\end{deluxetable}

\clearpage

\clearpage

\end{document}